\documentclass[12pt,preprint]{aastex}

\def\ga{\mathrel{\raise0.35ex\hbox{$\scriptstyle >$}\kern-0.6em
\lower0.40ex\hbox{{$\scriptstyle \sim$}}}}
\def\la{\mathrel{\raise0.35ex\hbox{$\scriptstyle <$}\kern-0.6em
\lower0.40ex\hbox{{$\scriptstyle \sim$}}}}
\def\co{CO {\it J}=1--0 }

\def\hij{high-{\it J} }
\def\loj{low-{\it J} }
 
\slugcomment{Accepted into ApJ}
 
\shorttitle{A ``blind'' detection of CO at $z \sim 2.5$}
\shortauthors{Lentati et al.}
 
\begin{document}

 \title{COLDz: Karl G. Jansky Very Large Array discovery of a gas-rich galaxy in COSMOS}

\author{L.~Lentati\altaffilmark{1},J. Wagg\altaffilmark{2,1}, C.~L. Carilli\altaffilmark{3,1}, D. Riechers\altaffilmark{4}, P. Capak\altaffilmark{5}, F. Walter\altaffilmark{6}, M. Aravena\altaffilmark{7}, E.~da Cunha\altaffilmark{6}, J.~A.~Hodge\altaffilmark{8}, R.~J.~Ivison\altaffilmark{9, 10}, I. Smail\altaffilmark{11}, C. Sharon\altaffilmark{4}, E. Daddi\altaffilmark{12}, R. Decarli\altaffilmark{6}, M. Dickinson\altaffilmark{13}, M. Sargent\altaffilmark{14}, N. Scoville\altaffilmark{5}, and V. Smol{\v c}i{\'c}\altaffilmark{15}}

\email{ltl21@mrao.cam.ac.uk}

\altaffiltext{1}{Astrophysics Group, Cavendish Laboratory, JJ Thomson Avenue, Cambridge CB3 0HE, UK}

\altaffiltext{2}{Square Kilometre Array Organisation, Jodrell Bank Observatory, Lower Withington, Macclesfield, Cheshire SK11 9DL, UK }

\altaffiltext{3}{National Radio Astronomy Observatory, Socorro, NM 87801, USA}

\altaffiltext{4}{Department of Astronomy, Cornell University, Ithaca, NY 14853, USA }

\altaffiltext{5}{California Institute of Technology, MC 105-24, 1200 East California Boulevard, Pasadena, CA 91125, USA}

\altaffiltext{6}{Max-Planck Institute for Astronomy, D-69117 Heidelberg, Germany}

\altaffiltext{7}{N\'{u}cleo de Astronom\'{\i}a, Facultad de Ingenier\'{\i}a, Universidad Diego Portales, Av. Ej\'ercito 441, Santiago, Chile}

\altaffiltext{8}{National Radio Astronomy Observatory, Charlottesville, VA 22903, USA}

\altaffiltext{9}{European Southern Observatory, Karl-Schwarzschild Strasse, 85748 Garching bei Munchen, Germany}

\altaffiltext{10}{Institute for Astronomy, University of Edinburgh, Blackford Hill, Edinburgh, UK}

\altaffiltext{11}{Institute for Computational Cosmology, Department of Physics, Durham University, South Road, Durham, DH1 3LE, UK}

\altaffiltext{12}{CEA-Saclay, Service d'Astrophysique, F-91191 Gif-sur-Yvette, France}

\altaffiltext{13}{National Optical Astronomy Observatory, 950 North Cherry Avenue, Tucson, AZ 85719, USA }

\altaffiltext{14}{Astronomy Centre, Department of Physics and Astronomy, University of Sussex, Brighton, BN1 9QH, UK}

\altaffiltext{15}{University of Zagreb, Bijeni\v{c}ka cesta 32, HR-10000 Zagreb, Croatia}

\begin{abstract}
The broad spectral bandwidth at mm and cm-wavelengths provided by the recent upgrades to the Karl G. Jansky Very Large Array (VLA) has made it possible to conduct unbiased searches for molecular CO line emission at redshifts, $z > 1.31$. We present the discovery of a gas-rich, star-forming galaxy at $z = 2.48$, through the detection of \co line emission in the COLDz survey, through a sensitive, Ka-band (31 to 39 GHz) VLA survey of a 6.5 square arcminute region of the COSMOS field. We argue that the broad line (FWHM$\sim$570$\pm$80 km s$^{-1}$) is most likely to be  \co at $z=2.48$, as the integrated emission is spatially coincident with an infrared-detected galaxy with a photometric redshift estimate of $z_{phot} = 3.2 \pm 0.4$. The \co line luminosity is $L'_{\rm CO} = (2.2\pm0.3) \times 10^{10}$ K km s$^{-1}$ pc$^2$, suggesting a cold molecular gas mass of M$_{ \rm gas} \sim (2 - 8)\times 10^{10}$ M$_{\odot}$ depending on the assumed value of the molecular gas mass to CO luminosity ratio $\alpha_{CO}$. The estimated infrared luminosity from the (rest-frame) far-infrared spectral energy distribution (SED)is $L_{\rm IR} = 2.5\times 10^{12}$ L$_{\odot}$ and the star-formation rate is $\sim$250 M$_{\odot}$ yr$^{-1}$, with the SED shape indicating substantial dust obscuration of the stellar light.  The infrared to CO line luminosity ratio is $\sim$114$\pm$19 L$_{\odot}$/(K km s$^{-1}$ pc$^2$), similar to galaxies with similar SFRs selected at UV/optical to radio wavelengths. This discovery confirms the potential for molecular emission line surveys as a route to study populations of gas-rich galaxies in the future.
\end{abstract}

\keywords{galaxies: star formation -- galaxies: high-redshift -- radio lines: galaxies}

\section{Introduction}

The evolution of the cosmic star formation rate density has now been studied as far back as 1~Gyr after the Big Bang, well into the epoch of reionization. Observational constraints on both the obscured and un-obscured star-formation in galaxies are now derived from sensitive multi-wavelength observations from the UV to long radio-wavelengths. However, our current understanding of the formation of galaxies is based almost exclusively on the stars, star formation, and ionized gas in those galaxies (e.g., Lyman-break selection, Steidel et al. 1996; magnitude-selected samples, e.g. Le Fevre et al. 2005; Lilly et al. 2007; narrow-band studies, e.g.\ Sobral et al.\ 2013). Nevertheless, it is now also possible to obtain a coherent and unbiased census of the dense, i.e. molecular gas (traced primarily by $^{12}$CO) that is fundamentally driving the history of star formation and stellar mass buildup in galaxies -- an important component that will take us towards a full physical understanding of the formation and evolution of galaxies throughout cosmic time. 
Until recently, most observations of the dense molecular CO-bright gas have been limited to targeted observations of FIR-luminous quasar host galaxies and submm-selected starburst galaxies (e.g. Bothwell et al.\ 2013; see review by Carilli \& Walter 2013). More recently, CO has been detected in lower luminosity systems at $z\sim 1.5$ (Daddi et al. 2008, 2010a; Tacconi et al. 2010, 2013; Genzel et al.\ 2012) which are claimed to be forming stars less efficiently than the far-infrared (FIR) luminous population. This has led to the claim that different star-formation laws should apply for starbursts and quiescent disk galaxies (e.g. Daddi et al. 2010b; Genzel et al.\ 2010; Sargent et al.\ 2013). Part of this may be due to the different CO-to-H$_2$ conversion factors adopted to estimate their cold molecular gas masses (e.g. Ivison et al.\ 2011; Riechers et al.\ 2011; Bolatto et al.\ 2013). 

To assess the gas properties of high-redshift galaxies, a ``blind'' survey is required. Past selection of IR luminous galaxies for follow-up CO line studies may have lead to a bias against cold, gas-rich systems. This selection effect may be overcome by conducting large volume surveys sensitive to \loj CO line emission.    
 The current generation of broad bandwidth mm and cm-wavelength facilities such as the the Karl G. Jansky Very Large Array (VLA), the Plateau de Bure Interferometer (PdBI) and the Atacama Large Millimeter/submillimeter Array (ALMA) now enable the possibility of conducting such surveys for molecular CO line emission over significant cosmological volumes.  Indeed, the first wide bandwidth survey of a 0.5 arcmin$^2$ region in the \textit{Hubble} Deep Field North across the 3mm band with the PdBI has revealed high-\textit{J} CO line emission in a few gas-rich galaxies (Walter et al.\ 2012, 2014; Decarli et al.\ 2014). These higher frequency surveys may be biased against the cold molecular gas traced by the lower-\textit{J} transitions of CO line emission that are accessible using the full 8~GHz of bandwidth now available with the VLA. Narrower bandwidth VLA and PdBI surveys toward galaxy overdensities (i.e. biased regions) at $z \sim 1 - 2$ have detected \loj CO line emission in gas-rich star-forming galaxies and AGN (Aravena et al.\ 2012; Wagg et al.\ 2012; Casasola et al.\ 2013; Tadaki et al.\ 2014).

Here, we present a ``blind'' VLA detection of CO~\textit{J}=1-0 line emission in a star-forming galaxy at $z=2.5$ in the COSMOS field. This is the first line emitter discovered in the CO luminosity density at high redshift (z) [COLDz] survey, a large survey programme (Riechers et al.\  \textit{in prep.}) aimed at providing an unbiased view of the evolution of molecular gas using the ground-state CO~\textit{J}=1-0 line, which is free from additional mass estimate uncertainties due to gas excitation, and the best tracer of the full size and mass of the gas reservoirs.
In \S 2 we describe the observations and data analysis, in \S 3 we present our analysis and results. We discuss our findings in \S 4 and conclude in \S 5. Throughout this work, we adopt a concordance $\Lambda$CDM cosmology throughout, with $H_0$ = 71~km~$\mathrm{s^{-1}Mpc^{-1}}$, $\Omega_{\mathrm{M}}$ = 0.27, and $\Omega_{\mathrm{\Lambda}}$ = 0.73 (e.g. Spergel et al.\ 2007).

\section{Observations and source identification}
\label{Section:Obs}

Ka-band observations of a $\sim$6.5 square arcminutes region of the  COSMOS extragalactic deep field (Scoville et al 2007a)  were made with the VLA in its D-configuration during the spring of 2013 (13A-398, PI: Riechers). This survey utilizes the full 8 GHz of bandwidth afforded by the VLA from 31-to-39 GHz, to simultaneously cover the peak epoch of galaxy assembly at $z\sim 2$ (in CO~\textit{J}=1--0; $\sim$ 3 Gyr after the Big Bang) and $z \sim 5.5$ (in CO~\textit{J}=2--1; $\sim$ 1 Gyr after the Big Bang), soon after the end of cosmic reionization.
 The full details of the survey design and data processing, along with an analysis of the complete spectral line and continuum data, will be presented in subsequent papers.  Here, we show a prominent example to demonstrate the effectiveness of the molecular emission line search technique for obtaining blind detections of CO line emission in high-redshift galaxies at long mm-wavelengths.

Observations were made using three frequency tunings of the VLA offset by 12 MHz, in order to account for the drops in sensitivity between each 128~MHz spectral window (\sc spw \normalfont ). J1041+0610 (S$_{34GHz}$  $\sim 0.65$ Jy) was observed as the phase calibrator, while 
3C286 was observed to calibrate the spectral bandpass and flux density scale.  
Calibration for each pointing and frequency setting was performed separately using a VLA data reduction pipeline developed using CASA by NRAO\footnote{https://science.nrao.edu/facilities/vla/data-processing/pipeline}, with 6 second time averaging performed prior to the analysis. We flag one 2~MHz wide channel at the upper and lower end of each \sc spw \normalfont so that combined with our frequency dithering strategy, we obtain nearly uniform sensitivity across the $\sim$8~GHz of spectral bandwidth. We also flag 30~MHz of spectrum between 31.48 and 31.51~GHz which is affected by strong radio frequency interference (RFI). We used natural weighting in the imaging and no cleaning was performed. To generate the continuum image, multi-frequency synthesis imaging was used and the resulting synthesized beamsize was $2\farcs 2 \times 1\farcs 9$.
Seven pointing mosaics of both the continuum and spectral line images were made for the full 8~GHz of total bandwidth between 31 and 39~GHz. The final rms in the spectral line data cube is $\sim$140~$\mu$Jy per 2~MHz channel and the mean of the synthesized beamsize (which varies over the band by $\sim$25\%) is $2\farcs 7\times 2\farcs 4$.  

In order to search for spectral line emitters in the COSMOS image cube, we apply a Bayesian source identification technique described further in \citealt{2014MNRAS.443.3741L},  to identify possible candidates. The COSMOS survey is a mosaic of several pointings, for which the level of noise changes as a function of position in the map, and this needs to be accounted for in order to assess the significance of a candidate.  To calculate the noise at any given position we use the sensitivity map for each channel, normalised to a peak of 1, calculate the rms at the center of each channel in the mosaiced image, and then scale that value according to the sensitivity map.  
Applying our line detection method to the VLA data, we detect with high significance a previously unknown line emitter at $\alpha = $10$^h$00$^m$18.21$^s$, $\delta = $02$^d$34$^m$56.7$^s$ (J2000), with an emission line frequency of 33.142~GHz and an integrated intensity of $0.076\pm0.010$~Jy~km~s$^{-1}$ (Figures \ref{figure:cont}-\ref{figure:5chan}). 
The linewidth is FWHM$\sim$570$\pm$80~km~s$^{-1}$. We hereafter refer to this source as COLDz J100018.21+023456.7. Our search algorithm also finds CO~\textit{J}=2--1 line emission in AzTEC-3 (Riechers et al.\ 2010a; Capak et al.\ 2011) amongst other sources, and these results will be presented in a future paper with an analysis of additional emission line candidates.

Our Ka-band VLA observations are complemented by a wealth of multi-wavelength data available in the COSMOS field (Scoville et al. 2007b), including deep \textit{Hubble} Space Telescope (HST) Advanced Camera for Surveys (ACS) I-band imaging; Subaru, Canada France Hawaii Telescope (CFHT), and Very Large Telescope Vista optical and near-IR imaging (McCracken et al.\ 2012); \textit{Spitzer} IRAC and MIPS 3.6-24$\mu$m imaging (Sanders et al 2007); \textit{Herschel} PACS 100 and 160~$\mu$m imaging and SPIRE 250, 350 and 500~$\mu$m imaging observations (Lutz et al.\ 2011; Oliver et al.\ 2012) and deep VLA 1.4~GHz imaging (Schinnerer et al.\ 2010). For a full description of these datasets see Ilbert\ et al. (2010, 2013). We also include an 850~$\mu$m upper-limit from SCUBA-2 (SCUBA-2 Cosmology Legacy Survey, private communication).

\section{Analysis and Results}
\label{Section:results}

\begin{figure*}
\begin{center}$
\begin{array}{c}
\includegraphics[width=130mm]{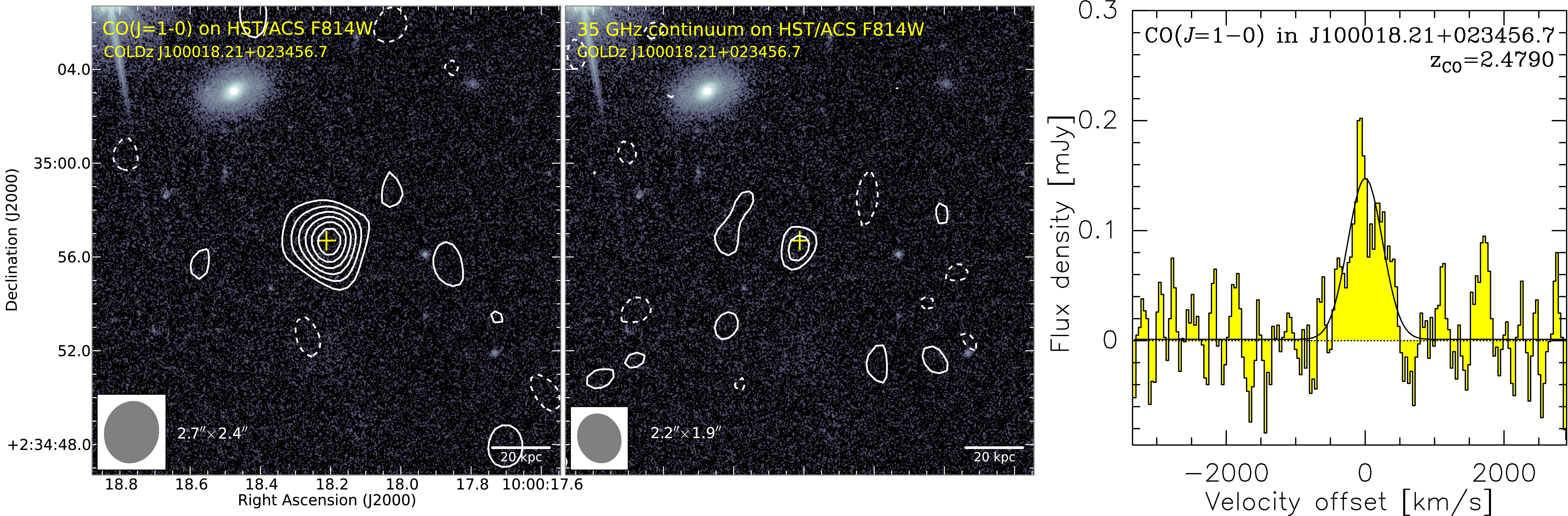} \\
\end{array}$
\end{center}
\caption{Left: Contours of the continuum subtracted moment 0 integrated intensity map overlaid on the {\it HST} ACS F814W image. The contour intervals are  (-2, 2, 3, 4, 5, 6, 7)$\times$0.01~Jy~km~s$^{-1}$.  
This ``blind'' detection has a significance of 7.8-$\sigma$ and is most likely to be \co at redshift 2.48. Center: Contour levels show the 35~GHz continuum emission at the location of the candidate source with contour intervals (-2, 2, 3)$\times 1.95$~$\mu$Jy per beam.  The continuum emission is detected  at the 3.5-$\sigma$ significance level. In both the left and centre plots dashed lines represent negative contours. Right: Spectrum of the \co line emission. This molecular gas-rich galaxy does not have an optical counterpart, suggesting a heavily obscured object similar to galaxies detected in mm-wavelength continuum surveys.}
\label{figure:cont}
\end{figure*}

\begin{figure*}
\begin{center}$
\begin{array}{c}
\includegraphics[width=130mm]{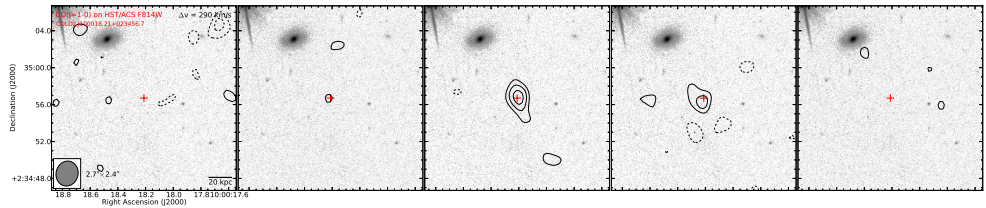} \\
\end{array}$
\end{center}
\caption{5 channel contour map overlaid on the {\it HST} ACS F814W image. In each panel the cross marks the same position as in Fig.\ref{figure:cont}, given by the center of a 2-dimensional Gaussian fit to the line emission.  Channels  are independent, and have been averaged to 32~MHz/290~km~s$^{-1}$, with an rms noise of 35$\mu$Jy~beam$^{-1}$. The frequency in the center panel is 33.146~GHz.  The contour intervals are  (-3, -2, 2, 3, 4, 5)$\times$35$\mu$Jy~beam$^{-1}$.  Dashed lines represent negative contours.  The ``blind'' detection has a significance of 2.3, 4.6 and 3.6 $\sigma$ across the center 3 panels.}
\label{figure:5chan}
\end{figure*}

Here, we describe the proposed multi-wavelength counterpart to the most significant, previously unknown line emitter identified in the COSMOS Ka-band spectral line data cube. The right panel of Fig.~\ref{figure:cont} shows the spectrum while the integrated CO line intensity map of J100018.21+023456.7 is shown in the left panel. The integrated line intensity is detected at 7.8-$\sigma$ significance, and corresponds to a redshift of either $z = 2.48$ for CO~\textit{J}=1-0, or $z = 5.96$ for CO~\textit{J}=2-1. The uncertainty on the redshift is 26~km~s$^{-1}$. 
We also find a tentative 3.5-$\sigma$ 35~GHz continuum counterpart with a flux density of $7 \pm 2 \mu$Jy at the same location (middle panel of Fig. \ref{figure:cont}).

Fig.~\ref{figure:5chan} shows a series of 5 contour maps overlaid on the {\it HST} ACS F814W image for the set of 32~MHz/290~km~s$^{-1}$ averaged channels that span the width of the source. The rms noise in the averaged channels is 35$\mu$Jy~beam$^{-1}$, and we find the ``blind'' detection has a significance of 2.3, 4.6 and 3.6 $\sigma$ across the center 3 panels.

\begin{table}
\caption{Infrared to cm-wavelength photometry of J100018.21+023456.7. Quoted uncertainties and upper-limits are $3\sigma$. SPIRE 250-500$\mu$m fluxes and uncertainties were extracted from the maps using 24$\mu$m priors, employing the same technique as described in Magnelli et al. 2012. Note that the formal confusion noise in each SPIRE band is $\sim$6mJy(Nguyen et al.\ 2010).}
\centering
\begin{tabular}{cc}
\hline\hline
wavelength  & flux   \\
		& $[\mu$Jy]  \\
\hline
7641\AA        &    $<$0.048 \\
2.2 $\mu$m & 2.19 $\pm$ 0.17\\
3.6 $\mu$m &  7.6 $\pm$   0.13\\
4.5 $\mu$m &  12.9  $\pm$ 0.23\\
5.8 $\mu$m &  32.7  $\pm$ 1.04\\
8.0 $\mu$m &  40.3  $\pm$ 2.55\\
24 $\mu$m &   153  $\pm$   11\\
165 $\mu$m &  $<$26130\\
250 $\mu$m &  16590 $\pm$  2000\\
350 $\mu$m &  19540 $\pm$  3400\\
500 $\mu$m &  11200 $\pm$  4200\\
850 $\mu$m &  $<$9000\\
8.8~mm &  7$\pm$2 \\ 
21~cm &  46$\pm$10  \\
\hline
\end{tabular}
\label{Table:SEDValues}
\end{table}

\begin{figure*}
\begin{center}$
\begin{array}{c}
\hspace{-1.5cm}
\includegraphics[width=180mm]{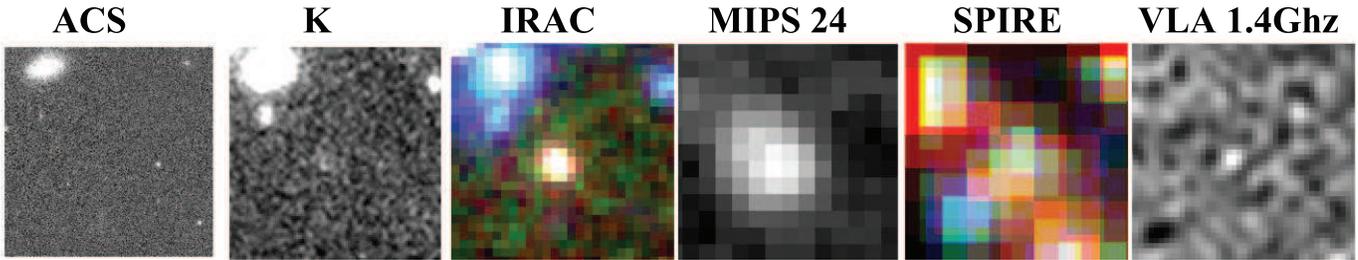} \\
\end{array}$
\end{center}
\caption{Multi-wavelength postage stamps centered on the peak of the line emission detected in J100018.21+023456.7. All postage stamps are $15'' \times 15''$ except for the SPIRE stamp, which is $90'' \times 90''$.
  Clear detections exist in the Spitzer 3-8$\mu$m IRAC and 24$\mu$m MIPS observations, as well as a colour composite image of the \textit{Herschel} SPIRE observations from 194 to 672$\mu$m. The multi-wavelength photometry suggests that this ultraluminous infrared galaxy likely has a significant mass in dust that is obscuring the optical counterpart.}
\label{figure:cutout}
\end{figure*}

\begin{figure}
\begin{center}$
\begin{array}{c}
\includegraphics[width=80mm]{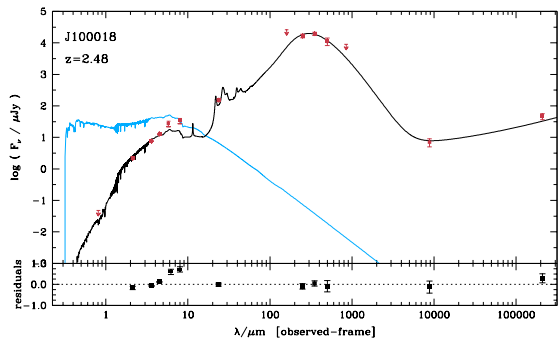} \\
\end{array}$
\end{center}
\caption{Spectral energy distribution for the COSMOS candidate source indicating that it is heavily dust obscured, forming stars at a rate of $\sim$250 M$_{\odot}$~yr$^{-1}$. We use a version of {\sc MAGPHYS} that accounts for extreme dust obscuration (da~Cunha et al.\ 2008; da~Cunha et al.\ \textit{in prep.}). Observed upper-limits are plotted at 3-$\sigma$ significance. Although the source is formally detected at 5.8 and 8.0$\mu$m, these points are considered to be upper-limits in the model fitting due to likely contamination by an AGN, which is not modeled in {\sc MAGPHYS}. The black line shows the best-fit {\sc MAGPHYS } model, while the blue line shows the corresponding dust-free (unextincted) stellar emission.  This source may be an analog of the very red far-infrared luminous sources reported in several other studies (e.g. Smail et al.\ 1999; Frayer et al.\ 2004; Aravena et al. 2010, Wang et al. 2010, Caputi et al. 2012, Walter et al. 2014; Decarli et al.\ 2014) that appear to make up a significant, but hard-to-study population at high redshift.}
\label{figure:sed}
\end{figure}

Fig. \ref{figure:cutout} shows multi-wavelength postage stamps at the position of the CO emission line candidate. These are, from left to right, the \textit{Hubble Space Telescope} ACS F814W,  K-band continuum, Spitzer IRAC and MIPS detections at 3-to-8~$\mu$m and 24 $\mu$m, \textit{Herschel} SPIRE observations from 250 to 500$\mu$m, and a VLA 1.4~GHz map.   Clear detections are obtained in the mid-infrared (MIR), the FIR and the radio.  
The flux density from the UV and infrared observations are given in Table \ref{Table:SEDValues} and plotted with the spectral energy distribution (SED) in Fig. \ref{figure:sed}. 
An initial fit to the spectral energy distribution of the counterparts to J100018.21+023456.7 leads to an estimate on the photometric redshift of $z_{phot} = 3.2 \pm 0.4$ \citep{2013A&A...556A..55I}, which corresponds within $2\sigma$ with the proposed CO redshift $z_{CO}=2.4790 \pm 0.0003$, suggesting this is CO~\textit{J}=1--0 line emission. The CO~\textit{J}=1-0 line luminosity is then $L'_{\rm CO} = (2.2 \pm 0.3) \times 10^{10}$K~km~s$^{-1}$~pc$^2$ with FWHM of $570\pm80$ km~s$^{-1}$. In order to estimate the total cold molecular gas mass, we must adopt a CO-to-H$_2$ conversion factor, $\alpha_{CO}$ (e.g. Bolatto et al.\ 2013). This value can be highly uncertain, even in the case of local luminous infrared galaxies (e.g. Papadopoulos et al.\ 2012), and typical values assumed for high-redshift galaxies are in the range, $\alpha_{CO} = (0.8 - 3.6)$~M$_{\odot}$ [K km~s$^{-1}$ pc$^2$]$^{-1}$ (e.g. Downes \& Solomon 1998; Daddi et al.\ 2010a; Tacconi et al.\ 2010; Bothwell et al.\ 2013). Assuming this range in values for the conversion factor, the estimated cold molecular gas mass for J100018.21+023456.7 is M$_{\mathrm{gas}}$ $\sim (2 - 8) \times 10^{10}$ ~M$_{\odot}$. 
Assuming $z = 2.48$, we use \sc MAGPHYS \normalfont (da~Cunha et al.\ 2008) with a prior on the star-formation history and dust attenuation that is appropriate for high-redshift galaxies and submillimeter galaxies (da~Cunha et al.\ \textit{in prep.}) to fit a model spectral energy distribution (SED) to the optical limits and rest-frame infrared through far-infrared photometry. With this Bayesian SED fitting algorithm (which includes two dust components with varying temperatures for assumed emissivity indices of $\beta = 1.5$ and 2.0) we obtain a luminosity-averaged dust temperature $\rm T_d = 40^{+2}_{-4}$~K, and a total dust mass of $\rm M_{d} = 4.3^{+1.6}_{-1.0}\times 10^8$~M$_{\odot}$. Adopting a gas-to-dust mass ratio of 100, typical of local ULIRGs (e.g. Santini et al.\ 2010), the estimated dust mass suggests a molecular gas mass that is consistent with the range estimated from the observed \co line luminosity and the assumed range in $\alpha_{CO}$. 
The best fit curve is plotted over the data points (Figure~\ref{figure:sed}), which results in an estimated stellar mass of $M_{*} \sim 2.5\times10^{11} $M$_{\odot}$ and infrared (8-1000$\mu$m)luminosity of L$_{\mathrm{IR}} = (2.5\pm 0.2)\times10^{12}$~L$_{\odot}$.  From the infrared luminosity we assume a Chabrier IMF \citep{2003PASP..115..763C} to estimate a star-formation rate of $\sim$250~M$_{\odot}$~yr$^{-1}$ (Kennicutt 1998), a specific star formation rate of $\sim 1$Gyr$^{-1}$, and a gas depletion timescale $\tau_{\mathrm{dep}} = 0.8-3.2 \times 10^8$yr. 
We note that the stellar mass is likely to be biased high due to the potential contamination of AGN emission to the restframe near-infrared (NIR) SED (Figure~\ref{figure:sed}). An upper-limit to the dynamical mass can be estimated from the CO linewidth and a limit on the size of the CO emission region ($<$2$''$) by assuming the ``isotropic virial estimator'' (Spitzer 1987). We estimate M$_{dyn} < 7.5 \times 10^{11}$~M$_{\odot}$, consistent with the molecular gas mass estimate and the upper-limit on the estimated stellar mass. 

\section{Discussion}
\label{Section:discussion}

The SED fit and subsequent estimate of the specific star-formation rate for  J100018.21+023456.7 suggests that it falls just below the scatter defined by the so-called `main sequence' of star-forming galaxies at z=2.5 (e.g. Elbaz et al.\ 2011), however we note that there is significant uncertainty associated with our stellar mass estimate for this very obscured source. 
It also appears to belong to a family of very red high-redshift ULIRGs reported in several studies (e.g. Smail et al.\ 1999; Frayer et al.\ 2004; Aravena et al. 2010; Wang et al. 2010; Caputi et al. 2012; Walter et al. 2014) that may make up a significant population at high redshift.  These galaxies have been found to be rich in gas and dust.

We estimate the infrared to CO line luminosity ratio ($L_{\rm IR}/ L'_{\rm CO}$) to be $\sim$114$\pm$19~L$_{\odot}$/(K~km~s$^{-1}$~pc$^2$), which is in good agreement with the median ratio measured for submm-selected star-forming galaxies (115$\pm$27~L$_{\odot}$/(K~km~s$^{-1}$~pc$^2$); Bothwell et al.\ 2013), and similar to the ratios observed in submm-faint, radio-selected galaxies (Chapman et al.\ 2008; Casey et al.\ 2009) and high-redshift galaxies selected at 24$\mu$m (Yan et al.\ 2010). The ratio in J100018.21+023456.7 is within the scatter, but higher than the mean ratio measured for BzK-selected star-forming galaxies at $z \sim 1-2$ (e.g. Daddi et al.\ 2010a; Tacconi et al.\ 2010), which typically have lower estimated star-formation rates. Comparing with less-luminous, gravitationally lensed star-forming galaxies (e.g. Baker et al.\ 2004; Coppin et al.\ 2007; Riechers et al.\ 2010b; Saintonge et al.\ 2013) the $L_{IR}/L'_{CO}$ ratios are typically $\sim$2$\times$ higher than that measured in J100018.21+023456.7, provided that these are not biased by differential magnification of the infrared and CO emission regions.  
It would therefore appear that this galaxy is similar to galaxies with similar SFRs selected at UV/optical to radio wavelengths.  The galaxy may fall on the main sequence, but it is more dust-obscured in the rest-frame UV/optical than typical galaxies on the main sequence at similar redshift. Thus, traditional selections for main sequence galaxies would
  not have identified this system.  Given the moderately deep limit at 850 $\mu$m and the derived  dust temperature, its properties are consistent with those found in
  single-dish 850 or 1100~$\mu$m surveys with typical dust temperatures of $T_d \sim$(30-40)~K (e.g. Kovacs et al.\  2006; Coppin et al.\ 2008; Swinbank et al.\ 2014), but in this case selected via CO line emission.  
Such galaxies could be discovered at shorter mm-wavelengths through their \hij CO line emission, and so a combination of VLA \co and higher-\textit{J} PdBI/ALMA 3~mm band spectral line surveys (e.g. Decarli et al.\ 2014; Walter et al.\ 2014), along with deep submm/mm-wavelength continuum observations may be sufficient to provide a {\it complete} survey of the luminous gas-rich star-forming population at $z \sim 2 -5$.  This will be further addressed by the forthcoming results of our analysis of the full Ka-band VLA survey.

\section{Conclusions}
\label{Section:Conclusions}

 Here, we present a ``blind'' detection of \co line emission at $z  = 2.48$ in COSMOS with the VLA Ka-band, the first such gas-rich galaxy discovered in our large VLA program. 
 We use a Bayesian search method which identifies the most significant emission line sources in the field, and present initial results from the search, providing details of a CO selected galaxy J100018.21+023456.7 with a likely redshift of $z = 2.48$. The estimated SFR is $\sim$250~M$_{\odot}$yr$^{-1}$, while the estimated cold molecular gas mass would be $M_{gas} \sim (2 - 8) \times 10^{10}$~M$_{\odot}$ for a plausible range in the CO-to-H$_2$ conversion factor. The $L_{\rm FIR}$-to-$L'_{\rm CO}$ ratio in J100018.21+023456.7 is similar to what is measured in other high-redshift star-forming galaxy populations detected in CO line emission.

Forthcoming publications of the COLDz survey will expand on the search results presented here, including VLA Ka-band observations of the GOODS-North field, applying alternative spectral line search methods.  These searches will provide constraints on the nature of further cold gas-rich galaxies in the both the COSMOS and GOODS-North data sets, allowing us to place new constraints on the cosmic density of cold molecular gas at $z \sim 2- 3$.

\acknowledgements We thank all those involved in the VLA project. These data were obtained as part of VLA observing program 13A-398 (PI: Riechers). 
The National Radio Astronomy Observatory is a facility of the National Science Foundation operated under cooperative agreement by Associated Universities, Inc. IRS acknowledges support from STFC (ST/L00075X/1), the ERC advanced investigator programme DUSTYGAL 321334 and a Royal Society/Wolfson Merit Award. RJI acknowledges support from the European Research Council (ERC) in the form of Advanced Grant, COSMICISM. DR, FW, EdC, JAH, IRS and VS acknowledge the Aspen Center for Physics and NSF Grant 1066293 for hospitality during the preparation of this study. VS acknowledges funding by the European Union's Seventh Frame-work program under grant agreement 337595 (ERC Starting Grant, 'CoSMass').


\end{document}